\documentclass[superscriptaddress,showpacs,dvips]{revtex4}
\usepackage{feynmp}
\usepackage{amssymb}
\usepackage{epsfig}
\usepackage{graphicx}
\usepackage{subfigure}
\usepackage{hyperref}

\begin{document}

\title{Competition between excitonic gap generation and disorder scattering in graphene}

\author{Guo-Zhu Liu}
\affiliation{Department of Modern Physics, University of Science and
Technology of China, Hefei, Anhui 230026, P. R. China}
\affiliation{Institut f$\ddot{u}$r Theoretische Physik, Freie
Universit$\ddot{a}$t Berlin, Arnimallee 14, D-14195 Berlin, Germany}
\author{Jing-Rong Wang}
\affiliation{Department of Modern Physics, University of Science and
Technology of China, Hefei, Anhui 230026, P. R. China}

\begin{abstract}
We study the disorder effect on the excitonic gap generation caused
by strong Coulomb interaction in graphene. By solving the
self-consistently coupled equations of dynamical fermion gap $m$ and
disorder scattering rate $\Gamma$, we found a critical line on the
plane of interaction strength $\lambda$ and disorder strength $g$.
The phase diagram is divided into two regions: in the region with
large $\lambda$ and small $g$, $m \neq 0$ and $\Gamma = 0$; in the
other region, $m = 0$ and $\Gamma \neq 0$ for nonzero $g$. In
particular, there is no coexistence of finite fermion gap and finite
scattering rate. These results imply a strong competition between
excitonic gap generation and disorder scattering. This conclusion
does not change when an additional contact four-fermion interaction
is included. For sufficiently large $\lambda$, the growing disorder
may drive a quantum phase transition from an excitonic insulator to
a metal.
\end{abstract}


\maketitle

%%%%%%%%%%%%%%%%%%%%%%%%%%%%%Main Body%%%%%%%%%%%%%%%%%%%%%%%%%%%%%%%%%%%%%

\section{Introduction}

The low-energy elementary excitations in undoped graphene are
massless Dirac fermions, which are believed to be responsible for
the unusual physical properties of this material \cite{CastroNeto,
DasSarma, Peres, Mucciolo, Ziegler, Kotov}. For clean undoped
graphene, the density of states (DOS) of fermions $N(\omega)$
vanishes linearly near the Dirac point, so the Coulomb interaction
between Dirac fermions is essentially unscreened \cite{CastroNeto,
Kotov, Gonzalez94}. When the Coulomb interaction is sufficiently
strong, $\lambda > \lambda_{c}$, it is able to open a finite gap for
the Dirac fermions by forming stable excitonic pairs. Once this
happens, the graphene then has an insulating ground state
\cite{Kotov, Khveshchenko00, Khveshchenko04, Gusynin, Herbut, Hands,
Drut09_1, Drut09_2, Liu, Gamayun}. This exotic insulator is induced
by the strong electron-electron interaction and can be classified as
a Mott insulator.

Although the excitonic insulating transition is a well-studied
problem in graphene, so far little work has been devoted to the
disorder effect on this nonperturbative phenomenon. In any realistic
graphene, there are always some kinds of disorders, which affect
significantly the low-energy transport properties of Dirac fermions.
In practice, the scattering of Dirac fermions by disorder can result
in important consequences \cite{Liu}. On the one hand, the disorder
scattering induces a finite density of states, $N(0)$, at the Fermi
level. The finite $N(0)$ leads to static screening of the Coulomb
interaction. Because the long-range nature of Coulomb interaction
plays an essential role in producing excitonic pairing \cite{Liu},
the fermion gap decreases with the growing disorder strength $g$ and
might finally close when $g$ is greater than some critical value
$g_c$. On the other hand, once the Dirac fermions acquire a finite
gap due to excitonic pairing, the scattering rate and low-energy DOS
will be quite different from those of gapless Dirac fermions.
Therefore, fermion gap generation and disorder scattering are not
independent. On the contrary, they have dramatic effects on each
other and thus should be treated within a unified formalism.
Technically, the fermion gap generation can be studied by the
Dyson-Schwinger gap equation approach while the disorder scattering
rate can be calculated by means of self-consistent Born
approximation (SCBA). In order to examine the disorder effect on
excitonic pairing, these two approaches should be combined properly.

In this paper, we examine the disorder effect on the excitonic
fermion gap generation caused by Coulomb interaction in graphene. By
solving the self-consistent equations for dynamical fermion gap $m$
and disorder scattering rate $\Gamma$, we found a critical line on
the plane of interaction strength $\lambda$ and disorder strength
$g$. This line separates two regions. In the region with large
$\lambda$ and small $g$, $m \neq 0$ and $\Gamma = 0$, so the ground
state is an excitonic insulator with zero dc electric conductivity.
In the other region, $m = 0$ and $\Gamma \neq 0$ for nonzero $g$, so
the Dirac fermions remain gapless and acquire a finite scattering
rate. In this region, the dc electric conductivity is $\sigma =
4e^2/\pi h$ to the leading order of Kubo formula, which is exactly
quantized and independent of disorder. Apparently, the excitonic
fermion gap generation due to Coulomb interaction competes strongly
with the fermion damping effect caused by disorder scattering. As a
consequence, the excitonic fermion gap and disorder scattering rate
can not be finite simultaneously in graphene. If we fix the
interaction parameter $\lambda$ at a sufficiently large value, then
the graphene may undergo a first-order phase transition from an
excitonic insulator to a metal at certain critical value $g_c$.
These results indicate that the excitonic insulating phase
transition can take place only in sufficiently clean graphene.

In addition to Coulomb interaction, there might be contact
four-fermion interaction in realistic graphene. Such interaction is
also able to generate a finite dynamical fermion gap when its
strength is sufficiently large. In this paper, we also include the
four-fermion interaction term in our self-consistent analysis and
find that it does not change the conclusion. In particular, the
excitonic fermion gap generation can not coexist with disorder
scattering even when additional four-fermion interaction is taken
into account.

The paper is organized as follows. In Sec.2, we study the
self-consistent equations of excitonic fermion gap and disorder
scattering rate when there is only Coulomb interaction. A strong
competition between excitonic fermion gap generation and disorder
scattering is found and a phase diagram on the $(\lambda,g)$ plane
is presented. In Sec.3, we study the self-consistent equations in
the presence of an additional contact four-fermion interaction. We
show that such interaction does not change our phase diagram. In
Sec.4, we discuss the physical implications of our result.

\section{Self-consistent analysis of excitonic transition and disorder scattering}

The Hamiltonian of massless Dirac fermion with Coulomb interaction
in graphene is given by
\begin{eqnarray}
H &=& v_{F}\sum_{\sigma=1}^{N}\int_{\mathbf{r}}
\bar{\psi}_{\sigma}(\mathbf{r})i\mathbf{\gamma}\cdot\mathbf{\nabla}
\psi_{\sigma}(\mathbf{r})\nonumber\\
&&+\frac{v_F}{4\pi}\sum_{\sigma,\sigma^{\prime}}^{N}
\int_{\mathbf{r},\mathbf{r}^{\prime}}
\bar{\psi}_{\sigma}(\mathbf{r})\gamma_0\psi_{\sigma}(\mathbf{r})
\frac{\lambda}{|\mathbf{r}-\mathbf{r}^{\prime}|}
\bar{\psi}_{\sigma^{\prime}}(\mathbf{r}^{\prime})\gamma_{0}
\psi_{\sigma^{\prime}}(\mathbf{r}^{\prime}),
\end{eqnarray}
where the effective interaction strength is defined as $\lambda=2\pi
e^2/v_F\epsilon$ with fermion velocity $v_F$ and dielectric constant
$\epsilon$. As usual, we adopt the four-component spinor field
$\psi$ to describe Dirac fermion and define the conjugate spinor
field as $\bar{\psi} = \psi^{\dagger} \gamma_{0}$. The $4\times 4$
$\gamma$-matrices satisfy the Clifford algebra \cite{Khveshchenko00,
*Khveshchenko04, Gusynin}. The fermion flavor is taken to be $N$ and
we will perform $1/N$ expansion since Coulomb interaction strength
may be too large to be an expansion parameter. The disorder
potential $U(\textbf{r})$ couples to fermion as
\begin{eqnarray}
H_\mathrm{dis} = \sum_{\sigma=1}^{N}v_{F}\int_{\mathbf{r}}
U(\textbf{r})\bar{\psi}_{\sigma}(\mathbf{r})\gamma_0
\psi_{\sigma}(\mathbf{r}),
\end{eqnarray}
where $\langle U(\textbf{r}) \rangle = 0$ and $\langle
U(\textbf{r})U(\textbf{r}^\prime)\rangle =
g\delta(\textbf{r}-\textbf{r}^\prime)$. Here, $U(\textbf{r})$ plays
the role of a random chemical potential and $g$ is the disorder
strength parameter. For notational convenience, in the following
discussion we will assume that $v_F = 1$ and restore $v_F$ whenever
necessary. The total Hamiltonian preserves a continuous chiral
symmetry $\psi \rightarrow e^{i\theta\gamma_5}\psi$, where
$\gamma_5$ anticommutes with $\gamma_{\mu}$, which will be
dynamically broken once a nonzero fermion gap is generated.

The free propagator of Dirac fermion is
\begin{eqnarray}
G_{0}^{-1}(i\omega_{n},\mathbf{k}) = i\omega_{n}\gamma_{0} -
\mathbf{\gamma}\cdot \mathbf{k},
\end{eqnarray}
where $\omega_n = (2n+1)\pi/\beta$ with $\beta = 1/T$ and $n$ being
integers. The Coulomb interaction and disorder scattering modify it
to the form
\begin{eqnarray}
G^{-1}(i\omega_{n},\mathbf{k}) = (i\omega_{n} +
i\mathrm{sgn}(\omega_{n})\Gamma)\gamma_{0} - \mathbf{\gamma}\cdot
\mathbf{k} - m,
\end{eqnarray}
where $\Gamma$ is the scattering rate due to interaction. The
fermion gap $m$ is generated by the excitonic pairing caused by
Coulomb interaction and can be obtained from the following
Dyson-Schwinger gap equation
\begin{eqnarray}
G^{-1}(p) &=& G_{0}^{-1}(p) + \int \frac{d^{3}k}{(2\pi)^{3}}
\gamma_{0}G(k)\gamma_{0} V(p-k),
\end{eqnarray}
to the leading order of $1/N$ expansion. The unscreened Coulomb
interaction is $V_{0}(q) = \frac{\lambda}{|\mathbf{q}|}$. After
including the dynamical screening effect from collective
particle-hole excitations, the effective interaction function has
the form
\begin{eqnarray}
V(q) = \frac{1}{V_{0}^{-1}(q)+\Pi(q)}.
\end{eqnarray}
The polarization function is defined as
\begin{eqnarray}
\Pi(q) = -N \int\frac{d^{3}k}{(2\pi)^{3}}\mathrm{Tr}
\left[\gamma_{0}G(k)\gamma_{0}G(k+q)\right].
\end{eqnarray}
It has the simple form
\begin{eqnarray}
\Pi(q) = \frac{N}{8} \frac{\mathbf{q}^{2}}{\sqrt{q_{0}^{2} +
|\mathbf{q}|^{2}}},
\end{eqnarray}
to the leading order of $1/N$ expansion. It vanishes linearly as
$\mathbf{q} \rightarrow 0$ in the static limit $q_{0}=0$, so the
Coulomb interaction remains long-ranged. For physical flavor $N=2$,
a sufficiently strong Coulomb interaction can trigger excitonic
pairing instability, which generates a dynamical gap for Dirac
fermions \cite{Kotov, Khveshchenko00, *Khveshchenko04, Gusynin,
Herbut, Hands, Drut09_1, *Drut09_2, Liu, Gamayun}.

When disorder scattering is included, a finite fermion damping rate
$\Gamma$ may be generated, which leads to two consequences. First,
it changes the energy spectrum of fermions. Second, it yields a
finite density of states at Fermi energy, which then screens the
long-range Coulomb interaction. Both these effects are important and
have to be incorporated into the gap equation. Because the sign of
$\Gamma$ is determined by frequency $\omega_n$, the gap equation
becomes much more complicated than that without $\Gamma$. To make
our theoretical analysis tractable, here we adopt the frequently
used instantaneous approximation \cite{Khveshchenko00, *Khveshchenko04, Gusynin}.
Taking trace on both sides of Eq.(5), we get
\begin{eqnarray}
m(i\varepsilon_n,\mathbf{p}) &=& \frac{1}{\beta}\sum_{\omega_n}\int
\frac{d^{2}\mathbf{k}}{(2\pi)^{2}}\frac{m(i\omega_n,\mathbf{k})}{(i\omega_n
+ i\Gamma\mathrm{sgn}(\omega_n))^2 + |\mathbf{k}|^2 +
m^2(i\omega_n,\mathbf{k})} \nonumber \\
&& \times V(i\varepsilon_n - i\omega_n,\mathbf{p} -
\mathbf{k},\Gamma).
\end{eqnarray}
It is possible to sum over $\omega_n$ only when the
frequency-dependence of fermion mass function is neglected, which is
the so-called instantaneous approximation \cite{Khveshchenko00,
*Khveshchenko04, Gusynin, Liu}. In this approximation, this equation
becomes
\begin{eqnarray}
m(\mathbf{p}) &=& \frac{1}{\beta}\sum_{\omega_n}\int
\frac{d^{2}\mathbf{k}}{(2\pi)^{2}} \frac{m(\mathbf{k})}{(i\omega_n +
i\Gamma\mathrm{sgn}(\omega_n))^2 + |\mathbf{k}|^2+
m^2(\mathbf{k})}\nonumber\\
&& \times V(\mathbf{p}-\mathbf{k},\Gamma).
\end{eqnarray}
After carrying out the frequency summation, we have
\begin{eqnarray}
m(\mathbf{p}) &=&\int \frac{d^{2}\mathbf{k}}{(2\pi)^{2}}
\frac{m(\mathbf{k})}{\sqrt{|\mathbf{k}|^2 +
m^2(\mathbf{k})}}\frac{1}{\pi}\nonumber\\
&&\times\mathrm{Im}\psi\left(\frac{1}{2}+\frac{\Gamma}{2\pi T} +
i\frac{\sqrt{|\mathbf{k}|^2 + m^2\left(\mathbf{k}\right)}}{2\pi
T}\right)V(\mathbf{p}-\mathbf{k},\Gamma).
\end{eqnarray}
To make an analytical computation, we take the limit $T\rightarrow
0$, so that
\begin{eqnarray}
\lim_{T\rightarrow 0}\mathrm{Im}\psi\left(\frac{1}{2} +
\frac{\Gamma}{2\pi T} + i\frac{\sqrt{\mathbf{k}^2 +
m^2\left(\mathbf{k}\right)}}{2\pi T}\right)
=\arctan\left(\frac{\sqrt{\mathbf{k}^2 +
m^2\left(\mathbf{k}\right)}}{\Gamma}\right),
\end{eqnarray}
which then leads to
\begin{eqnarray}
m(\mathbf{p}) &=& \int \frac{d^{2} \mathbf{k}}{(2
\pi)^2}\frac{m(\mathbf{k})}{\sqrt{|\mathbf{k}|^{2} +
m^{2}(\mathbf{k})}}
\frac{1}{\pi}\arctan\Big(\frac{\sqrt{|\mathbf{k}|^2 +
m^2(\mathbf{k})}}{\Gamma}\Big)\nonumber\\
&& \times \frac{1}{\frac{|\mathbf{p-k}|}{\lambda } +
\Pi(\mathbf{p-k},\Gamma)},
\end{eqnarray}
We now need an analytical expression for the polarization function
$\Pi(\mathbf{q},\Gamma)$, which is crucial to describe the screening
of long-range Coulomb interaction from particle-hole excitations. In
the Matsubara formalism, the polarization function within
instantaneous approximation is given by
\begin{eqnarray}
\Pi(\mathbf{q},T,\Gamma) = -\frac{N}{\beta}
\sum_{n=-\infty}^{+\infty}\int\frac{d^{2}\mathbf{k}}{(2\pi)^{2}}\mathrm{Tr}
\left[G_{0}(i\omega_{n},\mathbf{k})\gamma_{0}
G_{0}(i\omega_{n},\mathbf{k}+\mathbf{q})\gamma_{0}\right].
\end{eqnarray}
After introducing Feynman parameter and summing up imaginary
frequencies, the polarization function is found to have the form
\begin{eqnarray}
\Pi(\mathbf{q},T,\Gamma) &=& \frac{2N}{\pi^2}
\int_{0}^{1}dx\int_{C_{q}}^{\Lambda}dt\left\{\frac{C_{q}^2}{t^2}
\mathrm{Im}\left[\psi(\frac{1}{2} + \frac{\Gamma+it}{2\pi
T})\right]\right.\nonumber\\
&&\left.+\frac{t^2-C_{q}^{2}}{t}
\frac{\partial\mathrm{Im}\left[\psi(\frac{1}{2} +
\frac{\Gamma+it}{2\pi T})\right]}{\partial t}\right\},
\end{eqnarray}
where $C_{q}=\sqrt{x(1-x)}|\mathbf{q}|$ and an ultraviolet cutoff
$\Lambda$ is introduced. This expression is very complicated, but
can be further simplified in the zero temperature limit. As
$T\rightarrow 0$, we have $\lim_{T\rightarrow
0}\mathrm{Im}\left[\psi\left(\frac{1}{2} + \frac{\Gamma+it}{2\pi
T}\right)\right] = \arctan\left(\frac{t}{\Gamma}\right)$. After
tedious but straightforward computation, we finally have
\begin{eqnarray}
\Pi(\mathbf{q},\Gamma) &=& \frac{2N}{\pi^2}
\left[\Gamma\ln\left(\frac{\Lambda}{\Gamma}\right) +
F(|\mathbf{q}|,\Gamma)+|\mathbf{q}|\int_{0}^{1}dx \sqrt{x(1-x)}\right.\nonumber\\
&&\left.\times\arctan\Big(\frac{\sqrt{x(1-x)}|\mathbf{q}|}{\Gamma}\Big)\right],
\end{eqnarray}
where the function $F(x)$ is
\begin{eqnarray}
F(|\mathbf{q}|,\Gamma) &=& \left\{
\begin{array}{ll}
-\Gamma + \frac{\Gamma\sqrt{4\Gamma^2 -
|\mathbf{q}|^2}}{|\mathbf{q}|}
\arctan\left(\frac{|\mathbf{q}|}{\sqrt{4\Gamma^2 -
|\mathbf{q}|^2}}\right) & 0 < \frac{|\mathbf{q}|}{\Gamma} < 2
\\
-\Gamma -
\frac{\Gamma\sqrt{|\mathbf{q}|^2-4\Gamma^2}}{2|\mathbf{q}|}
\ln\left(\frac{|\mathbf{q}| - \sqrt{|\mathbf{q}|^2 - 4\Gamma^2}}
{|\mathbf{q}|+\sqrt{|\mathbf{q}|^2-4\Gamma^2}}\right) &
\frac{|\mathbf{q}|}{\Gamma} > 2
\\ -\Gamma. & \frac{|\mathbf{q}|}{\Gamma}=2
\end{array}
\right.
\end{eqnarray}
If we further let $|\mathbf{q}| = 0$, the polarization function
takes a finite value
\begin{equation}
\Pi(0,\Gamma) = \frac{2N}{\pi^2}\Gamma
\ln\left(\frac{\Lambda}{\Gamma}\right),
\end{equation}
which implies that the long-range Coulomb interaction now becomes
short-ranged. In fact, $\Pi(0,\Gamma)$ is precisely the Dirac
fermion DOS at zero energy $N(0)$ induced by disorder scattering
\cite{Durst}. As shown previously \cite{Liu}, the long-range nature
of Coulomb interaction plays an important role in driving excitonic
gap generation. Once the long-range Coulomb interaction is
statically screened by disorder scattering, excitonic gap generation
may be suppressed.

In the above equations, the scattering rate $\Gamma$ is simply
assumed to be an arbitrary constant. However, $\Gamma$ reflects the
fermion damping due to disorder scattering and may depends heavily
on the fermion gap. Indeed, fermion scattering rate $\Gamma$ and
fermion gap $m(\mathbf{p})$ can affect each other. In a more refined
treatment, $\Gamma$ should be calculated on the same footing with
fermion gap $m(\mathbf{p})$.

One frequently used approach of calculating $\Gamma$ is to first
average over random potential $U(\mathbf{r})$ and then make
mean-field analysis in the saddle point approximation, which amounts
to SCBA \cite{Fradkin, Lee93, Ando, Mirlin}. Although SCBA was
questioned recently \cite{Aleiner}, at present there is no better
choice when dealing with the interplay of disorder and excitonic
pairing. Within this approximation, the disorder scattering rate is
determined by the following equation
\begin{eqnarray}
\Gamma(\omega_n) = \frac{i}{4}\mathrm{Tr}\Big[\gamma_0
\int\frac{d^{2}\mathbf{k}}{(2\pi)^2} \frac{g}{(i\omega_{n} +
i\mathrm{sgn}(\omega_{n})\Gamma(\omega_n)) \gamma_{0} -
\mathbf{\gamma}\cdot \mathbf{k} - m(\mathbf{k})}\Big].
\end{eqnarray}
In principle, the scattering rate $\Gamma$ receives contributions
from both elastic scattering by quenched disorder and inelastic
scattering by Coulomb interaction. However, the later contribution
is forced to vanish at zero energy by the phase space restriction
originated from the Pauli exclusion principle. However, the
scattering rate from disorder can be finite even at zero energy.
Thus, we simply ignore the Coulomb interaction in this equation.

From the above equation, we know that $\Gamma(\omega_n)$ does not
depend on momentum $\mathbf{k}$ because any $\mathbf{k}$-dependence
will be lost after integration over $\mathbf{k}$. Further, no energy
is transferred during the disorder scattering process, so the
scattering rates $\Gamma(\omega_n)$ at different $\omega_n$ are
completely independent. Thus, we can drop the energy dependence of
$\Gamma(\omega_n)$ and focus only on the equation for $\Gamma(0)$.
In fact, in the derivation of gap equation, we have already assumed
that the scattering rate $\Gamma$ is independent of both momentum
and energy (otherwise it is impossible to get an analytical
expression for gap equation). Now we see that this assumption is
reasonable. Taking advantages of these simplifications, we
eventually have
\begin{equation}
1 = \frac{g}{2\pi}\int_{0}^{\Lambda} d|\mathbf{k}|
\frac{|\mathbf{k}|}{\Gamma^2 + |\mathbf{k}|^2 + m^2(\mathbf{k})}.
\end{equation}
When $m = 0$, the integration over $\mathbf{k}$ can be precisely
done yielding a scattering rate $\Gamma = \Lambda e^{-2\pi/g}$,
which has been known for many years \cite{Fradkin}. If we require
that the function $m(\mathbf{k})$ be given by fermion gap equation
Eq.(13), then we obtain two self-consistently coupled equations for
dynamical fermion gap and disorder scattering rate.

Before doing numerical computation, it is helpful to first make some
qualitative analysis. For simplicity, we assume a constant fermion
$m_0$. Now the integration over $\mathbf{k}$ in Eq.(20) can be
performed, yielding the solution
\begin{equation}
\Gamma^2 = \frac{\Lambda^2}{e^{4\pi/g}-1} - m_0^2.
\end{equation}
From this result, we know that $\Gamma$ can have a nonzero solution
only when $g$ is greater than some critical value
\begin{equation}
g > \frac{4\pi}{\ln(1+\frac{\Lambda^2}{m_0^2})}.
\end{equation}
Apparently, only sufficiently strong disorder can induce a finite
scattering rate when fermions are massive. From this qualitative
analysis, we expect a competition between excitonic gap generation
and disorder scattering.

\begin{figure}[ht]
\centering
   \includegraphics[width=2.5in]{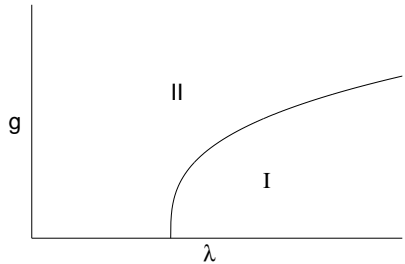}
\caption{Phase diagram of graphene on the plane of $(\lambda,g)$. In
region I, $m\neq 0$ and $\Gamma=0$. In region II, $m=0$ and $\Gamma
\neq 0$ when $g \neq 0$. This phase diagram is valid for weak
disorder.}
\end{figure}

In reality, the fermion gap $m(\mathbf{k})$ is not a constant, but
is a function of fermion momentum. It is therefore essential to
solve the equations (13) and (20) self-consistently. After carrying
out extensive numerical computation at physical flavor $N=2$, we
found a critical line on the plane of $(\lambda,g)$ and show the
phase diagram in Fig.1. In region I with large $\lambda$ and small
$g$, the dynamical fermion gap has a nontrivial solution, $m \neq
0$, but the disorder scattering rate have only trivial solution,
$\Gamma = 0$. In region II, the fermion gap has only trivial
solution, $m = 0$, while the disorder scattering rate acquires a
finite value, $\Gamma \neq 0$ (certainly, $\Gamma$ vanishes when
$g=0$). We did not find any evidence for the coexistence of
nontrivial $m$ and nontrivial $\Gamma$ in numerical computation. The
numerical results confirm our expectation that there is a
competition between the generation of excitonic fermion gap and the
generation of disorder scattering rate: fermion damping $\Gamma$
suppresses the formation of excitonic pairs, and fermion gap $m$
prevents the appearance of fermion damping.

In region I on Fig.1, the excitonic gap generation wins the
competition, hence the ground state is an excitonic insulator with
zero dc electric conductivity $\sigma = 0$. In region II, the
disorder scattering wins, then the ground state is gapless and the
Dirac fermions acquire a finite scattering rate. In region II, the
accurate calculation of electric conductivity is a problem in debate
\cite{CastroNeto, DasSarma}. To the leading order of the Kubo
formula, the dc electric conductivity is known to be $\sigma =
4e^2/\pi h$, which is independent of disorder and universal
\cite{Fradkin, Lee93, Ando, Mirlin, Durst}. When disorder strength
$g$ grows from I to II, $\sigma$ jumps suddenly from zero to a
universal value at certain critical value $g_c$. This is a
first-order excitonic insulator to metal phase transition driven by
growing disorder. Note, however, that the excitonic phase transition
is of first-order only in the presence of finite disorder, $g \neq
0$. In the clean limit, $g=0$, the excitonic transition is not of
first-order but is continuous as the Coulomb interaction parameter
$\lambda$ is varied. There is indeed a debate on the nature of this
transition in the clean graphene. An infinite-order transition was
claimed in some theoretical analysis \cite{Gusynin, Gamayun},
whereas a conventional second-order transition was found in recent
numerical work \cite{Drut09_1}.

As mentioned earlier, the disorder considered in this work is random
chemical potential, which may be generated by local defects, neutral
impurity atoms, or neutral adsorbed atoms in the plane of graphene
\cite{Peres, Mucciolo}. When this type of disorder is smooth at the
atomic scale, it will not mix the two inequivalent Dirac points. The
parameter $g$ is indeed the product of the concentration of impurity
atom (or defect) and the strength of a single impurity atom. In
practice, the magnitude of the critical disorder parameter $g_c$
depends on the fermion flavor and the Coulomb interaction strength
$\lambda$. If we take physical flavor $N=2$ and assume that graphene
is suspended in vacuum so that $e^2/v_F\epsilon \approx 2.16$
\cite{CastroNetoPhysics}, namely $\lambda\approx13.57$, then it is
easy to obtain $g_c \approx 0.36$. When graphene is placed on
certain substrate, the Coulomb interaction strength is reduced by
the screening due to substrate. From Fig. 1, we know that critical
value $g_c$ decreases as $\lambda$ is lowered.

We would like to emphasize the importance of making a
self-consistent analysis in our problem. In fact, if we solve the
fermion gap function (13) by assuming a finite constant scattering
rate $\Gamma$, then there is a critical value $\Gamma_c$. The gap
equation has no nontrivial solution for $\Gamma > \Gamma_c$, but a
finite fermion gap is opened for $\Gamma < \Gamma_c$. In this case,
there is coexistence of finite fermion gap and finite scattering
rate when $\Gamma < \Gamma_c$. Similarly, if we solve the SCBA
equation (20) by assuming a free constant gap $m_0$, there will be
coexistence of finite fermion gap and finite scattering rate when
the inequality (22) is satisfied. In both these cases, there will be
the third region with finite fermion gap $m$ and disorder scattering
rate $\Gamma$ lying between region I and region II on the phase
diagram. In this third region, the dc electric conductivity would be
\begin{equation}
\sigma = \frac{4e^2}{\pi h}\frac{\Gamma^2}{\Gamma^2 + m^2},
\end{equation}
which depends on disorder strength and displays metallic behavior
even when the fermions are gapped. However, when the equations (13)
and (20) are solved self-consistently, the fermion gap $m$ and
scattering rate $\Gamma$ can not be finite simultaneously, and can
not be zero simultaneously when $g \neq 0$. As a consequence, the dc
electric conductivity is either zero or exactly quantized, and does
not explicitly depend on disorder strength.

\section{Effect of additional four-fermion interaction}

In additional to Coulomb interaction, the contact four-fermion
interaction may also be important in realistic graphene. In the
language of field theory, such contact interaction is normally
described by the Gross-Neveu model \cite{Gross, Rosenstein}. It can
make additional contributions to the generation of finite excitonic
fermion gap \cite{Gross, Rosenstein}. A natural question is whether
the contact interaction alters the phase diagram shown in Fig.1.
When a four-fermion interaction is included in our self-consistent
analysis, it is in principle possible to obtain a coexistence of
finite excitonic fermion gap and finite scattering rate. In this
section, we will examine this possibility.

As an concrete example, we consider the following Gross-Neveu model
\begin{eqnarray}
H_{\mathrm{GN}} = \frac{G}{N}\sum_{\sigma = 1}^{N}\int
\left(\bar{\psi}_{\sigma}(\mathbf{r})\psi_{\sigma}(\mathbf{r})\right)^2.
\end{eqnarray}
This interaction term does not respect the continuous chiral
symmetry, but respects a discrete chiral symmetry. Its role in
excitonic gap generation was analyzed in recent years \cite{Herbut,
Hands, Liu, Gamayun}. Unlike Coulomb interaction, the interaction
strength does not depend on fermion momentum and energy and there is
no dynamical or static screening.

We first ignore the Coulomb interaction and consider the Gross-Neveu
model only. The corresponding gap equation is
\begin{eqnarray}
m(\mathbf{p}) &=& \frac{g'}{N\pi\Lambda}\int
\frac{d^{2}\mathbf{k}}{(2\pi)^{2}}
\frac{m(\mathbf{k})}{\sqrt{|\mathbf{k}|^{2} + m^{2}(\mathbf{k})}}\nonumber\\
&&\times\mathrm{Im}\psi\left(\frac{1}{2} + \frac{\Gamma}{2\pi T} +
i\frac{\sqrt{|\mathbf{k}|^2 + m^{2}(\mathbf{k})}}{2\pi T}\right).
\end{eqnarray}
where $g' = GN\Lambda$. From this equation, it is clear that fermion
gap is indeed independent of momentum. At zero temperature, we can
reduce gap equation to
\begin{eqnarray}
1 &=& \frac{g'}{2N\pi^2\Lambda}\int_{0}^{\Lambda}
d|\mathbf{k}||\mathbf{k}|\frac{1}{\sqrt{|\mathbf{k}|^{2} + m^{2}}}
\arctan\left(\frac{\sqrt{|\mathbf{k}|^2+m^2}}{\Gamma}\right).
\end{eqnarray}
In the clean limit, $\Gamma=0$, then
\begin{eqnarray}
1 &=& \frac{g'}{4N\pi\Lambda}\int_{0}^{\Lambda}
d|\mathbf{k}||\mathbf{k}|\frac{1}{\sqrt{|\mathbf{k}|^{2} + m^{2}}},
\end{eqnarray}
which has solution
\begin{eqnarray}
\frac{m}{\Lambda} =
\frac{1-\left(\frac{4N\pi}{g'}\right)^2}{\frac{8N\pi}{g'}}.
\end{eqnarray}
There is a critical coupling $g'_c = 4N\pi$. When $g' < g'_c$, there
is no non-trial solution for $m$; when $g' > g'_c$, there is
non-trivial solution for $m$. In the presence of disorder, the
scattering rate $\Gamma$ is given by the SCBA equation (20). Since
now fermion gap $m$ is a constant, it is easy to obtain the
following expression
\begin{equation}
\Gamma^2 = \frac{\Lambda^2}{e^{4\pi/g}-1} - m^2.
\end{equation}
After solving fermion gap equation (26) using this $\Gamma$, we
found no coexistence of finite fermion gap and finite scattering
rate.

We next consider the case when both Coulomb and Gross-Neveu
interactions are present in graphene. Recently, Gamayun \emph{et}
\emph{al.} showed that the analytical results can agree with
numerical simulation results once the Gross-Neveu model is included
\cite{Gamayun}. Now the whole gap equation has the form
\begin{eqnarray}
m(\mathbf{p}) &=&\int \frac{d^{2} \mathbf{k}}{4\pi^{2}}
\frac{m(\mathbf{k})}{\sqrt{|\mathbf{k}|^{2} +
m^{2}(\mathbf{k})}}\frac{1}{\frac{\mathbf{|p-k|}}{\lambda
}+\Pi(0,\mathbf{p-k},\Gamma)}\nonumber\\
&&\times\frac{1}{\pi}\arctan\left(\frac{\sqrt{|\mathbf{k}|^2
+ m^2(\mathbf{k})}}{\Gamma}\right)\nonumber \\
&& + \frac{g'}{N\Lambda}\int \frac{d^{2}\mathbf{k}}{4\pi^{2}}
\frac{m(\mathbf{k})}{\sqrt{|\mathbf{k}|^{2} +
m^{2}(\mathbf{k})}}\frac{1}{\pi}
\arctan\left(\frac{\sqrt{|\mathbf{k}|^2 +
m^2(\mathbf{k})}}{\Gamma}\right),
\end{eqnarray}
which couples self-consistently to SCBA equation (20). After
numerically solving these equations, we found no coexistence of
finite fermion gap and finite scattering rate. Although the
magnitude of fermion gap in the excitonic insulating phase becomes
larger after Gross-Neveu interaction is included, the qualitative
phase diagram shown in Fig.1 does not change.

\section{Summary and discussion}

In this paper, we studied the disorder effect on excitonic fermion
gap generation, i.e., excitonic insulating transition, due to
Coulomb interaction in graphene. By solving the self-consistent
equations of fermion gap and disorder scattering rate, we found a
strong competition between excitonic fermion gap generation and
disorder scattering. As a consequence of this competition, fermion
gap generation can not occur simultaneously with disorder scattering
in graphene. The phase diagram presented in Fig.(1) is the main new
output of this paper. We also showed that this phase diagram is not
changed by additional contact four-fermion interaction.

In any realistic graphene, the Dirac fermions are always scattered
by some kinds of disorder. Our results indicate that, even if the
Coulomb interaction is indeed sufficiently strong, the excitonic
transition will be completely suppressed if the disorder strength is
not small enough. Apparently, the excitonic transition can most
possibly be observed in very clean graphene.

We would point out that our results are obtained using SCBA. It
would be interesting to examine the effect of higher order
corrections. The most important corrections come from the
fluctuation effect associated with the Anderson localization. When
$\Gamma \neq 0$, the metallic electric conductivity $\sigma =
4e^2/\pi h$ is subjected to diffusion and Cooperon vertex
corrections \cite{Lee85}, which will lead to Anderson
metal-insulator transition. When Coulomb interaction is absent,
$\lambda = 0$, the localization of massless Dirac fermions was found
\cite{Aleiner, Altland}. For small but finite $\lambda$, though
Coulomb interaction is not strong enough to open fermion gap, it can
have important influence on transport properties \cite{Kotov,
Khveshchenko06, Mishchenko, Schmalian, Mishchenkoepl, Herbut2008,
Muller, Herbut2010}. In particular, it may destroy the fermion
localization \cite{Abrahams}. However, there is currently no widely
accepted theory for the interaction effect on localization
\cite{Abrahams}. On the experimental side, earlier measurements
suggested that the undoped graphene exhibits a universal minimum
conductivity \cite{CastroNeto}, $\sigma = 4e^2/h$. More recently, it
becomes clear from extensive transport experiments that the minimum
conductivity in undoped graphene is not universal but instead
strongly sample-dependent \cite{Mucciolo, Tan, ChenNP, ChenPRL}.
Regardless of the precise value of minimum conductivity, it seems
that the undoped, gapless graphene is metallic and free of
localization at experimentally accessible temperature
\cite{DasSarma, Peres, Mucciolo}, at least for weak disorder.

In this paper, we are mainly interested in the regime of strong
Coulomb interaction. For large $\lambda$, the localization effect is
even more involved because it is entangled with the non-perturbative
phenomenon of excitonic gap generation. As already explained, a
self-consistent treatment is crucial in our problem, hence excitonic
gap generation and fermion localization can not be studied
separately. Technically, the fermion gap function $m(\mathbf{p})$
appearing in (13) should be used when computing the diffusion and
Cooperon vertex corrections, while these vertex corrections should
be included in the polarization function $\Pi(\mathbf{p})$ and the
fermion gap equation (13), which correspond to the Altshuler-Aronov
type corrections \cite{Lee85}. Unfortunately, although in principle
it is possible to examine the importance of Anderson localization
effect, the self-consistent equations obtained in this way are too
complicated to be analyzed theoretically or numerically.

In this paper, we considered only one particular type of disorder,
i.e., random chemical potential. Our self-consistent analysis may be
extended to study the effects of other types of disorders
\cite{Ludwig}, such as random gauge field or random mass, on
excitonic pairing formation. Specifically, ripples are believed by
many people to be important in graphene and thus attracted intensive
investigation \cite{Guinea}. Such ripple configuration can be
described by a random gauge potential \cite{Guinea}. It is
interesting to study the ripple effect and to examine whether
ripples can drive an analogous phase transition in the future.

We would like to thank P. Pyatkovskiy for pointing out a mistake in
our previous calculation of polarization function and W. Li and J.
Wang for helpful discussion. G.Z.L. is supported by the National
Science Foundation of China under Grant No.11074234 and the project
sponsored by the Overseas Academic Training Funds of University of
Science and Technology of China.

\end{document}